\documentstyle[fullpage,epsf,amssymb]{article}

\def\eg		{{\it eg.,~}}
\def\etal	{{\it et al.~}}
\def\etc	{{\it etc.}}
\def\ie		{{\it i.e.,~}}

\def\12 {\frac{1}{2}}

\begin{document}
\bibliographystyle{plain}

\centerline{\Large Fitting random stable solar systems to Titius-Bode laws}

\medskip

\centerline{Preprint: 10 Oct. 1997}
\vskip 0.1in
\centerline{\sc \large Wayne Hayes}
\centerline{Dept. of Computer Science, 10 King's College Rd.,}
\centerline{University of Toronto, Toronto, Ontario, M5S 3G4, Canada}
\centerline{wayne@cs.utoronto.ca}
\vskip 0.1in
\centerline{\sc \large Scott Tremaine}
\centerline{Canadian Institute for Theoretical Astrophysics and}
\centerline{Canadian Institute for Advanced Research Program in Cosmology and
    Gravity,}
\centerline{McLennan Labs, University of Toronto,}
\centerline{60~St.\ George St., Toronto M5S 3H8, Canada}
\centerline{tremaine@cita.utoronto.ca}
\vskip 0.1in
\vskip 0.1in

\begin{abstract}
Simple ``solar systems'' are generated with planetary orbital radii $r$
distributed uniformly random in $\log r$ between 0.2 and 50 AU. A
conservative stability criterion is imposed by requiring that adjacent
planets are separated by a minimum distance of $k$ Hill radii, for
values of $k$ ranging from 1 to 8.  Least-squares fits of these systems
to generalized Bode laws are performed, and compared to the fit of our
own Solar System.  We find that this stability criterion, and other
``radius-exclusion'' laws, generally produce approximately
geometrically spaced planets that fit a Titius-Bode law about as well
as our own Solar System.  We then allow the random systems the same
exceptions that have historically been applied to our own Solar
System.  Namely, one gap may be inserted, similar to the gap between
Mars and Jupiter, and up to 3 planets may be ``ignored'', similar to
how some forms of Bode's law ignore Mercury, Neptune, and Pluto.  With
these particular exceptions, we find that our Solar System fits
significantly better than the random ones.  However, we believe that
this choice of exceptions, designed specifically to give our own Solar
System a better fit, gives it an unfair advantage that would be lost if
other exception rules were used.  We conclude that the significance of
Bode's law is simply that stable planetary systems tend to be regularly
spaced; this conclusion could be strengthened by the use of more
stringent methods of rejecting unstable solar systems, such as
long-term orbit integrations.
\end{abstract}

\begin{flushright}
{\small
``For a statistician, fitting a three-parameter curve of uncertain form
to ten points \\
with three exceptions certainly brings one to the far
edge of the known world.'' \\
--- Bradley Efron (1971)}
\end{flushright}

\section{Introduction}

The Titius-Bode ``law'',
\begin{equation}
r_i = 0.4 + 0.15 \times 2^i, \;\;\;\; i=-\infty, 1,\ldots, 8, 
\label{eq:bodelaw}
\end{equation}
roughly describes the planetary semi-major axes in astronomical units,
with Mercury assigned $i=-\infty$, Venus $i=1$, Earth $i=2$, \etc~
Usually the asteroid belt is counted as $i=4$. The law fits the planets
Venus through Uranus quite well, and successfully predicted the
existence and locations of Uranus and the asteroids.  However, (i) the
law breaks down badly for Neptune and Pluto; (ii) there is no reason
why Mercury should have $i=-\infty$ rather than $i=0$, except 
that it fits better that way; (iii) the total mass of the
asteroid belt is far smaller than the mass of any planet, so it is not
clear that it should be counted as one. The question of the
significance of Bode's law has taken on increased interest with
discoveries of extra-solar planets, and is also worth
re-examination because computer speeds now permit more powerful
statistical tests than were previously possible.

A fairly comprehensive history of the law and attempts to explain it up
to the year 1971 can be found in Nieto (1972)\nocite{Nieto72}.  Most
modern arguments concerning the validity of Bode's law can be assigned
to one of three broad classes:

\begin{enumerate}

\item  Attempts to elucidate the physical processes leading to Bode's
law.  These are based on a variety of mechanisms, including dynamical
instabilities in the protoplanetary disk (Graner \& Dubrulle 1994;
Dubrulle \& Graner 1994; Li \etal
1995)\nocite{LiZhangLi95,GranerDubrulle94,DubrulleGraner94},
gravitational interactions between planetesimals (Lecar
1973)\nocite{Lecar73}, or long-term instabilities of the planetary
orbits (Hills 1970; Llibre \& Pi\~nol 1987; Conway \& Elsner
1988)\nocite{Hills70,LlibrePinol87,ConwayElsner88}.  We shall not
comment on these explanations, except to say that we find none of them
entirely convincing.

\item Discussions that ignore physics but try to assess whether the
success of Bode's law is statistically significant.  Good (1969)
\nocite{Good69} performs a likelihood test under the null hypothesis
that the planet distances should be distributed uniformly random in
$\log r$.  He includes the asteroid belt, but ignores Mercury, Neptune,
and Pluto, subjectively assigning (\ie guessing) a factor of 5 penalty
to his likelihood ratio for ignoring these planets.  He concludes that
there is a likelihood ratio of 300--700 in favour of Bode's
law being ``real'' rather than artifactual.  Efron (1971)
\nocite{Efron71} attacks Good's analysis, in particular his choice of
null hypothesis (Good's and Efron's articles are followed by over a
dozen extended ``comments'' from other statisticians).  Efron notes
that the difference between semi-major axes of adjacent planets is an
increasing function of distance for all adjacent planet pairs except
Neptune-Pluto.  He proposes, without physical basis, that this {\it law
of increasing differences} is a better null hypothesis, the only reason
cited being that Bode's law ``predicts'' increasing differences.
Duplicating Good's analysis with this new null hypothesis, he computes
a likelihood ratio in favour of Bode's law of only 8:5, and concludes
that ``there is no compelling evidence for believing that Bode's law is
not artifactual.''  Conway \& Zelenka (1988)\nocite{ConwayZelenka88}
repeat Efron's analysis using the law of increasing differences, this
time ignoring only Pluto, and computing a more realistic penalty for
doing so.  They compute a likelihood ratio of approximately unity, also
concluding that Bode's law is artifactual.  We believe these analyses
are flawed because there is no physical basis for the law of increasing
differences; in fact later we will show that systems that are stable
according to our criteria only rarely satisfy the law of increasing
differences.

\item Discussions of other laws that may influence the spacing of
the planets.  Many of these involve resonances between the mean motions
of the planets, such as Molchanov (1968; but see H\'enon
1969)\nocite{Molchanov68,Henon69}, Birn (1973)\nocite{Birn73}, and
Patterson (1987)\nocite{Patterson87}. A promising development is the
recognition that planets are capable of migrating significant distances
after their formation (Fern\'andez \& Ip 1984; Wetherill 1988; Ipatov
1993; Lin, Bodenheimer \& Richardson
1996)\nocite{FernandezIp84,Wetherill88,Ipatov93,LinBodenheimerRichardson96}.
For example, this process can explain the resonant relationship between
Neptune and Pluto (Malhotra 1993, 1995)\nocite{Malhotra93,Malhotra95}
and may explain the spacing of the terrestrial planets (Laskar
1997)\nocite{Laskar97}.

\end{enumerate}

The present paper combines the first two of these approaches: we
generate a broad range of possible model solar systems and exclude
those that are known to be dynamically unstable.  We then ask which of
the remaining ones satisfy laws similar to Bode's.

\section{Method}

\subsection{Radius-exclusion laws}

A necessary, but not sufficient, condition for the stability of a solar
system is that its planets never get ``too close to each other'' (Lecar
1973)\nocite{Lecar73}.  This can be formalized into several
``radius-exclusion laws''.  
\begin{enumerate}
\item Simple scaling arguments for near-circular, coplanar orbits suggest
that a test particle on a stable orbit
cannot approach a planet more closely than $k$ Hill radii for some $k$,
using the Hill radius $h$ as defined by Lissauer (1987,
1993)\nocite{Lissauer87,Lissauer93},
\begin{equation}
h = H_M r,\;\;\;\;\; H_M = \left(\frac{M}{3M_\odot}\right)^{\frac{1}{3}}
\label{eq:HillRadius}
\end{equation}
for a planet of mass $M$, semi-major axis $r$, and {\it fractional Hill
radius} $H_M$.  We shall extend this criterion to two adjacent planets
with non-zero mass by summing their respective Hill radii.  There are
other plausible ways to combine adjacent planets: it might be more
physically reasonable to use the sum of the masses to define a single
combined Hill radius, although it is not clear that this is preferred
when more than two planets are involved.  Exponents other than $1/3$
may also be reasonable (Wisdom 1980; Chambers \etal
1996)\nocite{Wisdom80,ChambersWetherillBoss96}.  However, the
difference between these approaches is probably unimportant given the
uncertainty in $k$, as discussed below.
\item For non-circular orbits, we also expect that the aphelion distance
of the inner planet is less than the perihelion distance of the outer one.
In other words, if the $i^{\rm th}$ planet has semi-major axis $r_i$ and
eccentricity $e_i$, we expect that $ r_i(1+e_i) < r_{i+1}(1-e_{i+1})$.
Taking this further, we may demand that the planets are separated by a
Hill radius even at their closest possible approach, giving
$$ r_i(1+e_i + H_{M_i}) < r_{i+1}(1 - e_{i+1} - H_{M_{i+1}}).$$
\item Several authors have argued that boundaries between stable and
unstable orbits occur at resonances of the form $j:(j+1)$ (Birn 1973;
Wisdom 1980; Weidenschilling \& Davis 1985; Patterson 1987; Holman \&
Murray 1996).
\nocite{Birn73,Wisdom80,WeidenschillingDavis85,Patterson87,HolmanMurray96}
Weidenschilling \& Davis (1985)\nocite{WeidenschillingDavis85} argue
that two planets are unlikely to form closer than their mutual 2:3
resonance (because small solid bodies are trapped in the outer
$j:(j+1)$ resonances of a protoplanet due to gas-induced drag; once
trapped, their eccentricities are pumped up, causing orbit crossing).
For the 2:3 resonance, we can define $H_{2:3}$ by using Kepler's
third law to define $R_{2:3} = (3/2)^{2/3}$, and splitting the
distance between two adjacent planets by solving $R_{2:3} = 1 +
H_{2:3} + R_{2:3} H_{2:3}$.
\end{enumerate}
Combining all three of
Hill radii, eccentricities, and the 2:3 resonance, we obtain
$$ r_i (1 + V_i) < r_{i+1} (1 - V_{i+1}),
    \;\;\;\;\;\;\; V_i = \max(H_{2:3}, e_i + H_{M_i}).$$

Clearly our results will be highly dependent upon the extent of radius
exclusion.  For the Hill radius of equation (\ref{eq:HillRadius}),
which was derived for the case of two small planets orbiting a
massive central object, a value of $k$=2--4 is believed to leave the
two planets in permanently stable orbits (Wetherill \& Cox 1984, 1985;
Lissauer 1987; Wetherill 1988; Gladman
1993)\nocite{WetherillCox84,WetherillCox85,Lissauer87,Wetherill88,Gladman93}.
For more than two planets, recent work by Chambers \etal
(1996)\nocite{ChambersWetherillBoss96} suggests that 
{\em no} value of $k$ gives permanent stability.  Instead, the
stability timescale grows exponentially with increasing orbit
separation, with billion-year stability for our solar system requiring
$k\gtrsim 13$.  Furthermore, simulations of the stability of test
particles in the current Solar System (Holman
1997)\nocite{Holman97Kuiper} seem to show that there remain few stable
orbits in the outer Solar System other than those near Trojan points.  This
provides circumstantial evidence that a small value of $k$ is
not enough to separate stable orbits, since the outer planets are
separated from each other by more than 15 Hill radii.
For these reasons, our experiments use several radius-exclusion laws,
including various combinations of Hill radii, 2:3 resonances,
eccentricities, and $k$. 

It is easy to see why radius-exclusion laws tend to produce planetary
distances that approximately follow a geometric progression.  If a fixed
fractional radius exclusion $V$ is used for every planet, and planets
are packed as tightly as possible according the radius-exclusion law,
then the physical extent of radius exclusion at distance $r$ is $rV$,
and the resulting planetary separations would follow an exact geometric
progression with semi-major axis ratio $(1+V)/(1-V)$.  If the planets
are packed less tightly, then the progression will be only approximately
geometric.

\subsection{Generating and fitting solar systems}

We generate 9 planet distances $r_i, i=0,\ldots,8$, distributed uniformly
random in $\log r$ between 0.2 and 50 AU, constrained so that the exclusion
radii of adjacent planets do not overlap. We generated 4096 samples for each
of the radius-exclusion laws listed in Table \ref{tab:Hills}.
\begin{table}
\begin{tabular}{|l|l|l||l|l|}
\hline
Name & Description & trials & Excl $\Rightarrow$ LID & LID $\Rightarrow$ Excl\\
\hline
$M_0$ & Each planet has zero mass & 1 & 0.4\% & 100\%\\
$1H_i$ & Hill radii corresponding to planets of our Solar System & 1.92 & 0.8\% & 93.3\%\\
$2H_i$ & Like $1H_i$, except radius exclusion of 2 Hill radii & 3.94 & 1.6\% & 87.2\%\\
$M_J$ & All planets have Jupiter's fractional Hill radius & 7.88 & 1.9\% & 49.3\%\\
$4H_i$ & Like $1H_i$, except radius exclusion of 4 Hill radii & 19.3 & 4.5\% & 57.6\%\\
$H_i+e_i$ & Adjacent planets no closer than eccentricity + 1 Hill radius & 44.1 & 3.4\% & 23.1\% \\
$H_{2:3}$ & Adjacent planets no closer than the 2:3 resonance & 230 & 10.5\% & 24.8\%\\
$\max$ & Adjacent planets no closer than $\max(H_i+e_i, H_{2:3})$& 479 & 9.7\% & 16.9\% \\
$8H_i$ & Like $1H_i$, except radius exclusion of 8 Hill radii & 2820 & 11.7\% & 1.7\%\\
\hline
\end{tabular}
\caption{The various radius-exclusion laws we used. The table is
ordered by the ``trials'' column, which is the 4096-sample average
number of Monte-Carlo trials required from a $\log$-uniform
distribution to find a sample that satisfies the corresponding radius
exclusion criterion.  The
last two columns (see \S\ref{sec:LID}) list the percentage of
samples for which the law of increasing
differences (LID) agrees with exclusion laws for a 9-planet system.
For example, every sample satisfies $M_0$ exclusion, but only 0.4\% of
those satisfy LID.}
\label{tab:Hills}
\end{table}
For $e_i$, we use the maximum eccentricity for each planet over the
past 3 million years (Quinn \etal 1991)\nocite{QuinnTremaineDuncan91}.
Table \ref{tab:Hills} lists the average number of Monte-Carlo trials
required to find a list of distances satisfying the radius-exclusion
law.  If a list of distances did not satisfy the exclusion criterion,
the entire list was discarded.  The mean number of trials needed to build a
``valid'' system that satisfies the radius-exclusion criterion
increases as the exclusion radii get larger.

For each sample that satisfies the relevant radius-exclusion, we perform
a nonlinear least-squares fit of the distances $r_i$ to
\begin{equation}
a + b c^i,
\label{eq:abc}
\end{equation}
which we call a ``generalized Bode law''.  The fit is performed 
by minimizing the objective function
\begin{equation}
\chi^2=\sum_{i=0}^8 \left(\frac{\log(a+b c^i) - \log r_i}{\sigma_i}\right)^2,
\label{eq:objective}
\end{equation}
constrained so that $a,b > 0$ and $c > 1$.
We fit on $\log r$ rather than $r$ because we want the fractional
error of each planet to be weighted equally.
In the Appendix, we derive the standard deviations $\sigma_i$ analytically.
The initial guess for the parameters in the objective function is 
$$ c_0 = \frac{1}{8} \sum_{i=0}^{7} \frac{r_{i+1}}{r_i},$$
$$ b_0 = r_8 / c_0^8, $$
$$ a_0 = \max(0, r_0 - b_0). $$

To more accurately reflect the various forms of Bode's law, we also
attempt fits that ``ignore'' 1, 2, and 3 planets.  We do this by
performing fits on all ${9 \choose j},\; j=1,2,3$, possible
combinations of ignoring $j$ out of $9$ planets, and choosing the best
fit for each $j$.  At this point we have 5 ``fits'' for each solar
system: an initial guess, and a fit that ignores 0, 1, 2, and 3
planets.  We then repeat the entire procedure, allowing one gap to be
inserted between the two adjacent planets with the largest
$r_{i+1}/r_i$ ratio, to mimic the gap between Mars and Jupiter.  This
gives us ten fits out of $2 \sum_{j=0}^3 {9\choose j} = 260$ trials for
each solar system.

\section{Results}

\subsection{Fits}

\begin{figure} [hbt]
\epsfxsize=14cm
\epsfysize=10cm
\leavevmode
\centering
\epsffile{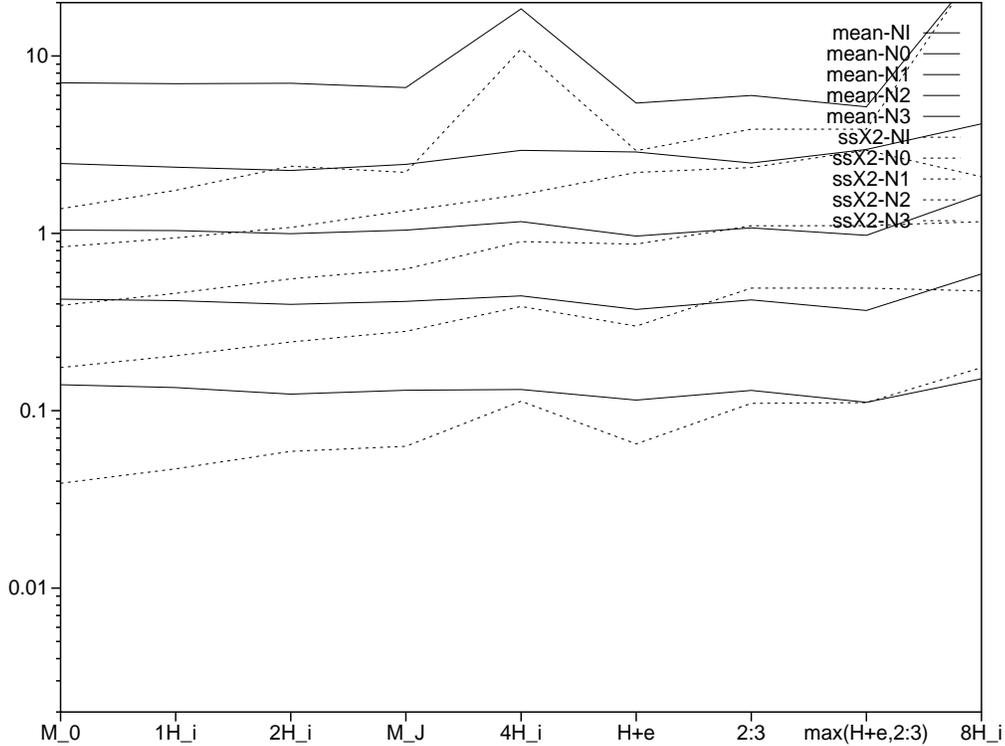}
\caption{$\chi^2$ fits, using equation (\ref{eq:objective}),
when no gap between planets (analogous to the asteroid
belt) is allowed (label `N'). The $\chi^2$ of our own Solar System is
shown by dashed lines, and the means of 4096 random ones are shown by
solid lines. The labels on the horizontal axis correspond to the first
column of Table \ref{tab:Hills}.  The labels I, 0, 1, 2, and 3 denote
the fit for the initial guess, and 0, 1, 2, 3 planets ignored,
respectively.}
\label{fig:X2-N}
\end{figure}

\begin{figure} [hbt]
\epsfxsize=14cm
\epsfysize=10cm
\leavevmode
\centering
\epsffile{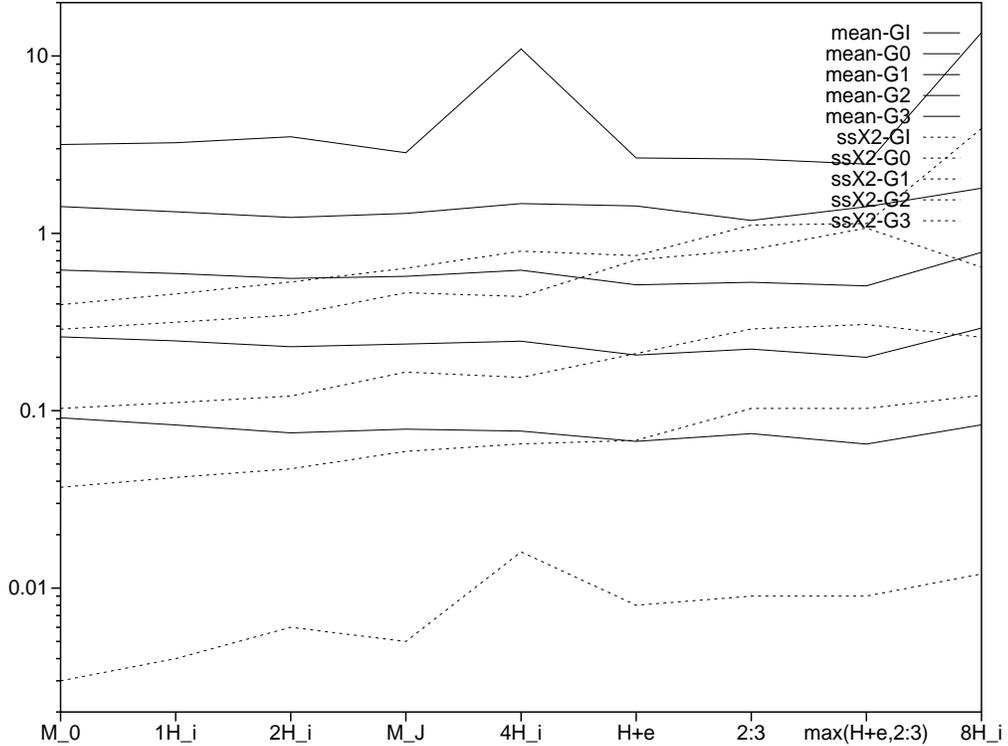}
\caption{Same as Figure \ref{fig:X2-N},
but with a gap allowed (label `G').}
\label{fig:X2-G}
\end{figure}

\begin{figure} [hbt]
\epsfxsize=14cm
\epsfysize=10cm
\leavevmode
\centering
\epsffile{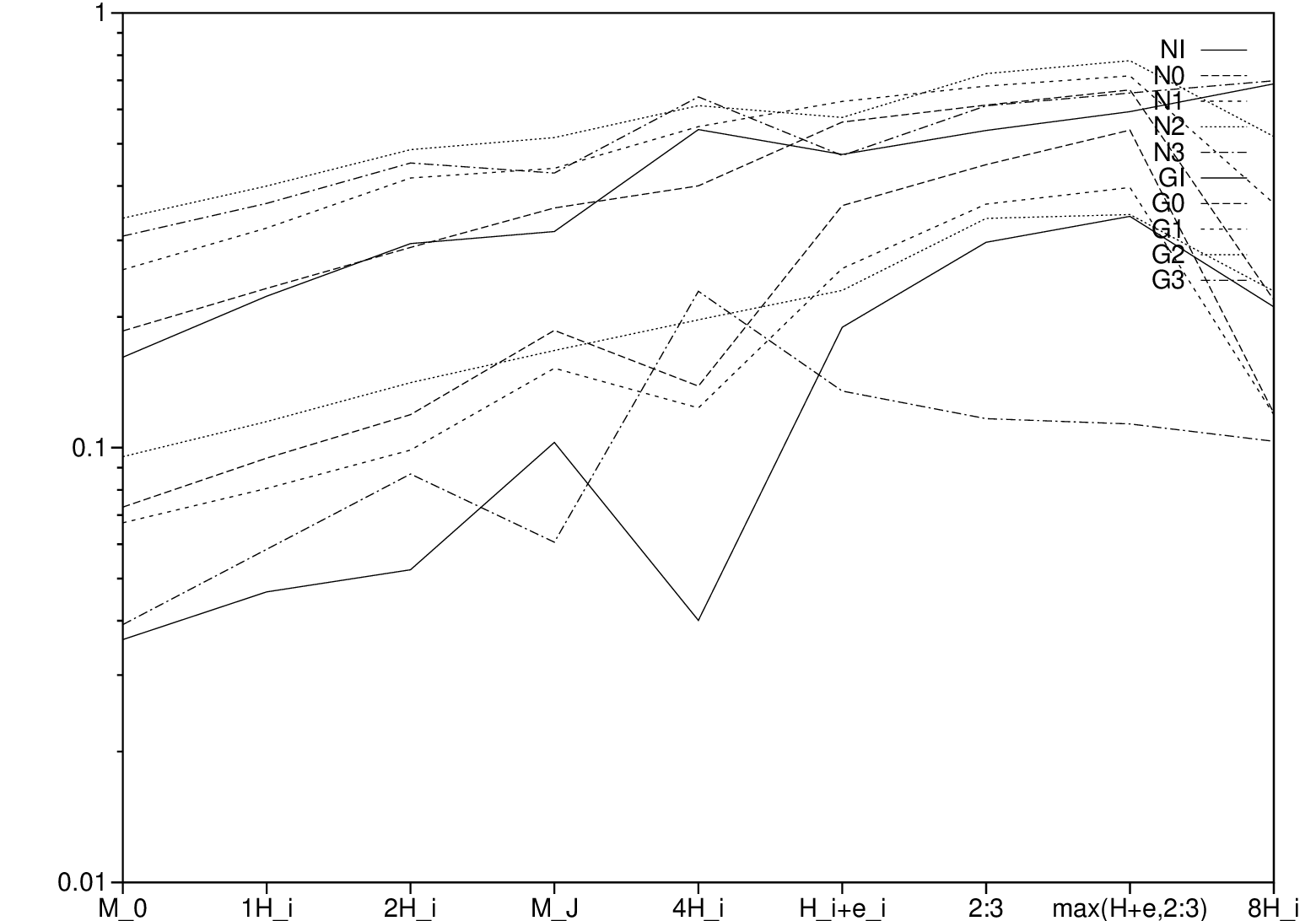}
\caption{The quantile of the Solar System's $\chi^2$, compared to the
4096 random ones, including systems both allowing a gap (`G'), and not
(`N').  The labels are the same as in Figure \ref{fig:X2-N}.
}
\label{fig:ssQu}
\end{figure}

Results of all the fits for each type of system are presented in Figures
\ref{fig:X2-N}, \ref{fig:X2-G}, and \ref{fig:ssQu}.  Not surprisingly, the
best fit for our own Solar System occurs when a gap is added between Mars and
Jupiter, while Mercury, Neptune, and Pluto are ignored.  Even for our own
Solar System, the best fit depends slightly on the radius-exclusion law, since
this affects the denominator $\sigma_i$ (see Appendix); for the
three cases in which the radius exclusion for each planet is identical
in $\log r$ ($M_0, M_J,$ 2:3), the fit for our Solar System is identically
$$ 0.450 + 0.132 \times 2.032 ^ i, \;\;\;\; i=0,1,\ldots,8,$$
with $\chi^2$ values of 0.003, 0.005, and 0.009, respectively. This
result can be compared to the original Bode's law, equation
(\ref{eq:bodelaw}).

Consider the case where no gaps are allowed, and up to 3 planets may be
ignored (Figure \ref{fig:X2-N}).  Note first that the values of
$\chi^2$ for the initial guess (top solid curve) are roughly equal to
6, which is the expected value since there are 6 degrees of freedom (9
planets minus the 3 parameters of Bode's law). The only exceptions are
the cases $4H_i$ and $8H_i$; this is probably because the $\sigma_i$ in
equation (\ref{eq:objective}) are only approximate when the radius
exclusions are different for each planet (as is the case for all the
$kH_i$ systems, $k=1,2,4,8$), and the approximation worsens as the
radius exclusion increases.

When there is no radius-exclusion law (left edge of the Figure), the
$\chi^2$ for the Solar System is always substantially less (by a factor
3--6) than the mean of the random systems with the same number of
ignored planets; this is consistent with the conclusion that the Solar
System satisfies a generalized Bode law significantly better than
random systems. However, as we apply more stringent radius-exclusion
laws (moving right in the Figure), the $\chi^2$ values for the Solar
System become quite similar to the mean of the random systems,
indicating that the Solar System is no closer to a generalized Bode law
than random systems. This situation changes in Figure \ref{fig:X2-G},
which shows the case where a gap is allowed. For every radius-exclusion
law, our Solar System's best $\chi^2$ value with three planets ignored
and one gap (bottom dashed line in Figure \ref{fig:X2-G}) is 1--1.5
orders of magnitude smaller than the mean of the random systems with
the same number of exceptions.  We suggest that this is because the
particular exceptions we investigated were historically designed
specifically to make our Solar System fit better.

Figure \ref{fig:ssQu} shows the Solar System's quantile---the fraction
of random systems with the same exceptions that have a $\chi^2$ smaller
than the Solar System. The quantile is not exceptional if no gap is
allowed (upper five curves), ranging from about 0.15 to 0.7, and only
mildly exceptional if a gap is allowed ($\gtrsim0.04$), and generally
becomes less exceptional as the random solar systems are chosen with
increasingly stringent radius-exclusion laws (moving right in the
Figure).

\begin{figure} [hbt]
\epsfxsize=14cm
\epsfysize=10cm
\leavevmode
\centering
\epsffile{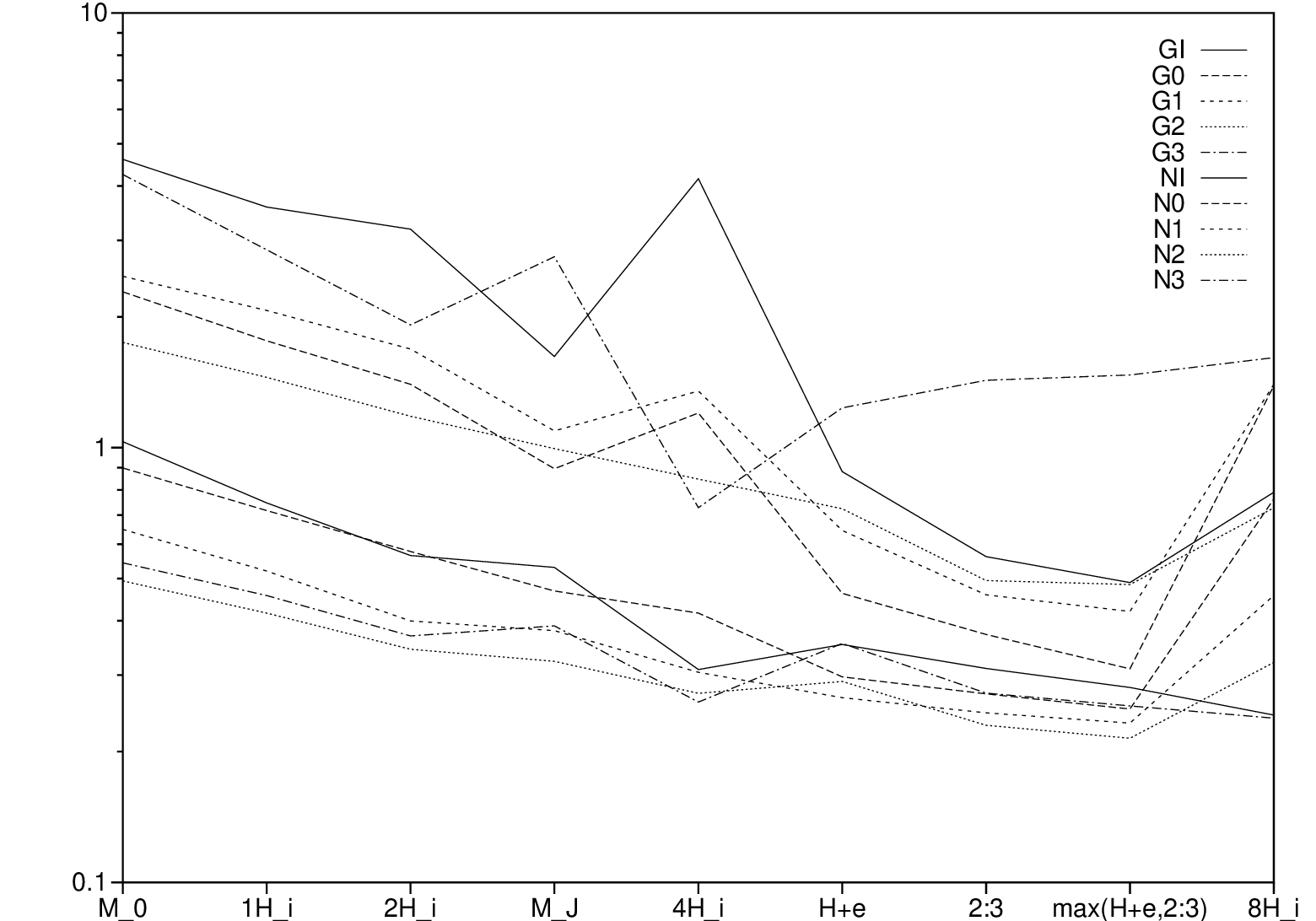}
\caption{The ``likelihood ratio'' of Bode's law being ``real'', using
the formula of Good, Efron, and Conway \& Zelenka (eq. \ref{eq:likelihood}).  
The labels are the same as in Figure \ref{fig:ssQu}.
}
\label{fig:ssLkli}
\end{figure}

To compare our results to those of Good (1969)\nocite{Good69}, Efron
(1971)\nocite{Efron71}, and Conway \& Zelenka
(1988)\nocite{ConwayZelenka88}, we have used the same---though rather
arbitrary---procedure advocated by Efron: we compute the ``likelihood
ratio'' $L$ of Bode's law being ``real'', as opposed to artifactual,
using the formula
\begin{equation}
L = {1\over 4Q}\left(1-{\textstyle{1\over3}}\right)
\label{eq:likelihood}
\end{equation}
where $Q$ is the Solar System's quantile, the $(1-\frac{1}{3})$ factor
is meant to penalize us for restricting ourselves to a geometric law
(as opposed, say, to one based on $n^2$), and the $\frac{1}{4}$ factor
is designed to normalize the ratio so that anything within about one
standard deviation of the mean has a likelihood ratio less than 1.  The
results of this calculation are presented in Figure \ref{fig:ssLkli}.
The highest likelihood ratios occur for the $M_0$ case, since the
random systems are least constrained when there is no radius exclusion,
but even these are all $\lesssim 5$, which is not particularly
significant.

\begin{figure} [hbt]
\epsfxsize=14cm
\epsfysize=10cm
\leavevmode
\centering
\epsffile{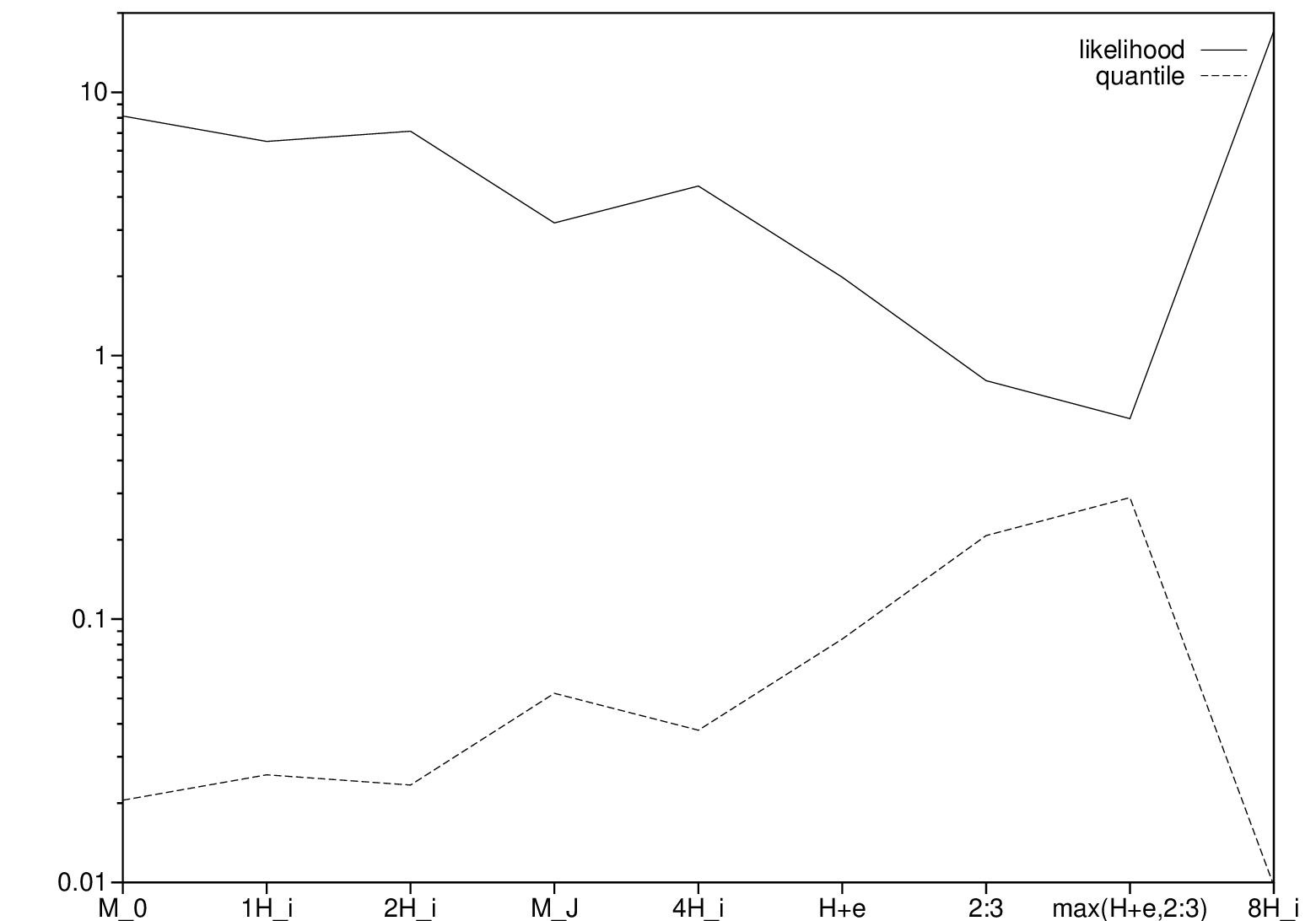}
\caption{The quantile and likelihood ratio of the Solar System,
compared to 4096 random ones, when no exceptions are allowed, but
the main asteroid belt is included as a ``planet'' at radius 2.8 AU.}
\label{fig:N10ssQu}
\end{figure}

One could also argue that the main asteroid belt should ``count'' as an
object. In Figure \ref{fig:N10ssQu}, we display the Solar System's
quantile and likelihood ratio for this case, with no exceptions
allowed.  In this case, the Solar System's $\chi^2$ value is generally
in the top 2 to 10 percentile, with a likelihood ranging from unity up
to 8.  (We ignore the $8H_i$ case, which is probably badly skewed due
to the approximate nature of the $\sigma_i$.)  This likelihood ratio is
also not particularly significant, given that the case we are examining
is, to some extent, tailored to the properties of our Solar System.

The value of the parameter $a$ in equation (\ref{eq:abc}) is zero in
about 25\% of the samples, representing an exact geometric progression,
while the remainder distribute uniformly up to about 0.5 AU.  The value
of $b$ is roughly bimodal, with peaks near 0 and 0.2;
the first tends to occur when $a$ is non-zero and the second when $a$
is near zero. The value of $c$ clusters around 1.5--1.8 when no planets
are ignored, and slightly higher if some are.  There was no observable
correlation of $a,b,$ or $c$ with $\chi^2$.

\begin{figure} [hbt]
\epsfxsize=14cm
\epsfysize=10cm
\leavevmode
\centering
\epsffile{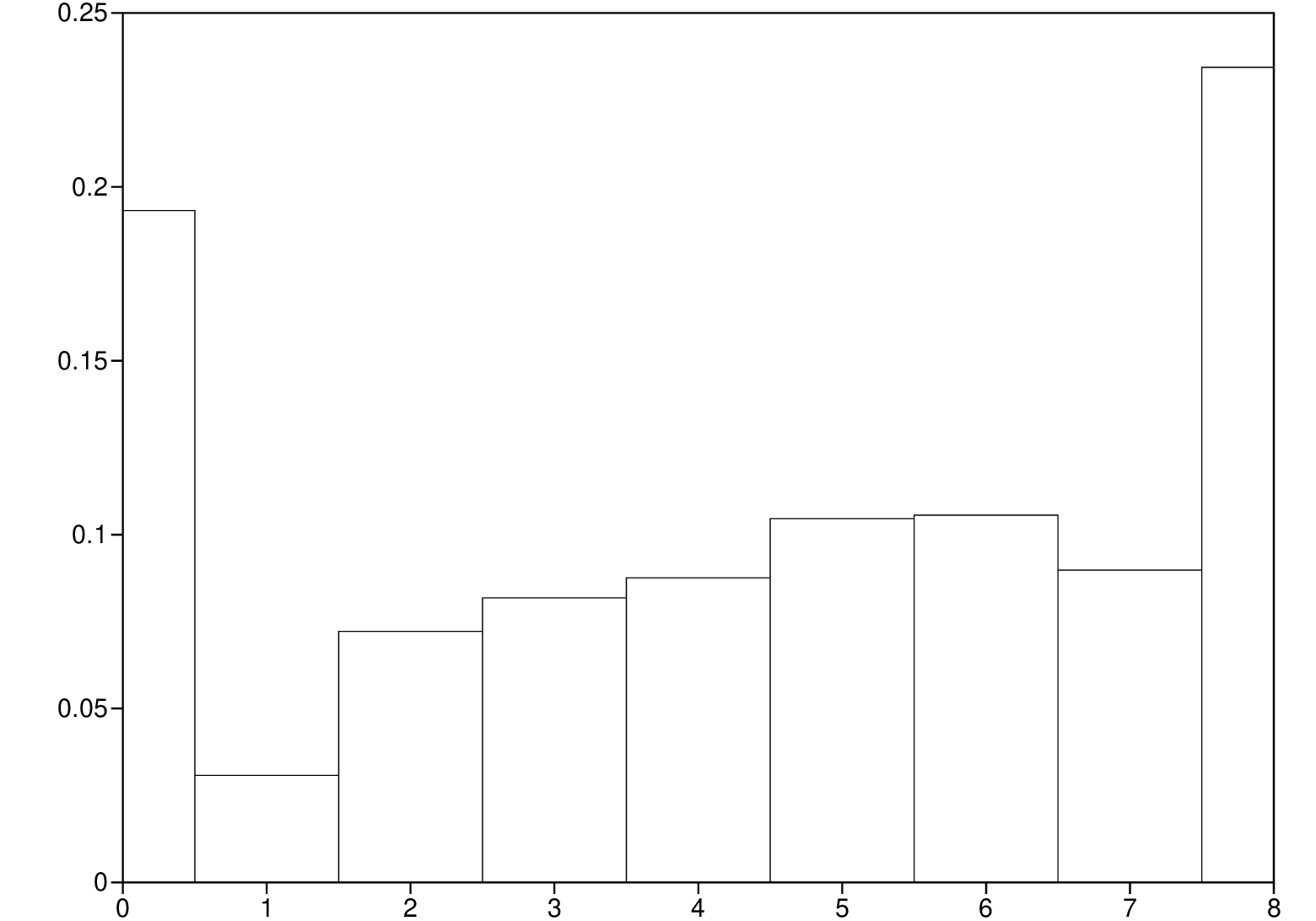}
\caption{A histogram showing how often each planet is chosen to be
ignored, as a fraction of the total, when ignoring one planet in the
$M_0$ system.  The distribution does not change significantly in other
systems.}
\label{fig:ignoreHist}
\end{figure}

A histogram of which planet gets ignored in the $(M_0,N1)$ systems is
shown in Figure \ref{fig:ignoreHist}.  The innermost and outermost
planets are ignored most often, which is expected since a planet with
only one neighbor is less constrained than those with two.
The placement of the gap is approximately uniform between all planet
pairs, for all types of systems.  Adding a gap was observed to produce
a better fit in about 65\% of all cases.

It is prudent to show that our results are not strongly dependent upon the
assumption that the underlying distribution is uniform in $\log r$. 
If instead we assume a disk surface density that scales as $r^{-3/2}$,
which roughly corresponds to the expected density in the protoplanetary
disk (Lissauer 1993)\nocite{Lissauer93}, then if all the planets are
equally massive, they should be uniformly distributed in $\sqrt{r}$.  We
therefore performed our entire suite of experiments again, this time
trying to fit an $a+b c^i$ law to solar systems with an underlying
random distribution that is uniform in $\sqrt{r}$.  We find that most
of the above results are qualitatively unchanged.  For example, in the
$N0$ case, the fit of the random systems worsens, so that our Solar
System's quantile moves down, but only by about 0.05, to 0.15.
In the $G3$ case, the Solar System's quantile is almost
the same in the $\sqrt r$ distribution as in the $\log r$
distribution.  Furthermore, as radius exclusion increases, the effect
of the underlying distribution is suppressed because radius exclusion
is biased towards accepting solar systems that follow a roughly
geometric progression.
We conclude that our comparisons are not strongly affected by the
assumption that the underlying distribution is uniform in $\log r$.

\subsection{The law of increasing differences}
\label{sec:LID}
Efron (1971)\nocite{Efron71} and Conway \& Zelenka
(1988)\nocite{ConwayZelenka88} have noted that the distance between
planets in the Solar System is an increasing function of distance for
all adjacent pairs except Neptune-Pluto.  They propose that this is a
reasonable law on which to base a null hypothesis concerning the
statistical significance of Bode's law.  They note that a pure
$\log$-uniform distribution produces increasing differences only a
small percentage of the time, and that if we assume the law of
increasing differences then the success of Bode's law is unsurprising.

We are uncomfortable with this law because it has no physical basis.
This concern has prompted us to examine the relation between
radius-exclusion laws and the law of increasing differences. Table
\ref{tab:Hills} shows the occurrences of agreement between the law of
increasing differences, and radius exclusions.
As Table \ref{tab:Hills} shows, (i) a
system that satisfies radius exclusion rarely satisfies the law of
increasing differences; (ii) one that satisfies the law of increasing
differences will often satisfy all but the most stringent exclusion
laws.  We also observed that
(iii) the number of trials required to find a random
$\log$-uniform sample that satisfies the law of increasing differences
is 1 to 2 orders of magnitude larger than that for radius exclusion;
(iv) random solar systems that satisfy the law of increasing
differences have $\chi^2$ values that are 1.3 to 3 times smaller than
ones generated using radius exclusion.  Thus, the law of increasing
differences is a much more restrictive assumption than radius-exclusion
laws, and in the absence of any physical justification, it does not
form a sound basis from which to judge the validity of Bode's law.

\section{Discussion and conclusions}

We have measured the deviation from Bode's law of solar systems whose
planetary distances are distributed uniformly random in $\log r$,
subject to radius exclusion constraints.  We find that, as radius
exclusion becomes more stringent, the systems tend to fit Bode's law
better.  We compare these fits to that of our own Solar System.  We
find that, when no exceptions or gaps are allowed, our Solar System
fits marginally better than random systems that follow weak
radius-exclusion laws, but fits no better, or even worse, than those
that satisfy more stringent but still reasonable radius exclusions.  If
one gap is allowed to be added, and up to 3 planets are ignored, then
our Solar System fits significantly better than random ones built with
weak radius exclusions, and marginally better than ones with strong
radius exclusions; however, this modest success for Bode's law probably
arises because these rules (3 planets removed, 1 gap added) were
designed specifically in earlier centuries to make our Solar System fit
better.

Even though our underlying distance distribution, uniform in $\log r$,
is scale invariant, the analysis of Graner and Dubrulle (1994) and
Dubrulle and Graner (1994) does not apply to most of our cases, since
the distribution of planetary masses and Hill radii is not
scale-invariant. We found that, even in the cases where the
radius-exclusion law {\em is} scale-invariant ($M_0$, $M_J$, 2:3),
the best-fitting generalized Bode law has $a\not=0$ in 36\% of the
cases, and hence is {\em not} scale-invariant.

Our approach to Bode's law is very simplistic. We ignore all of the
details of planet formation, and use simple stability criteria to
discard unstable systems.
The natural next step is to replace the approximate stability criteria
we have used by actual orbit integrations.  We conjecture that if we
repeat the experiments of the present paper using a direct integration
of several Gyr (\eg Wisdom \& Holman 1991\nocite{WisdomHolman91}) to
discard unstable systems, we would find that the surviving systems
fit a generalized Bode law better than
the ones in this paper and
approximately as well as our own Solar
system, showing that the significance of Bode's law is simply that
stable planetary systems tend to be regularly spaced.

\vskip 0.4cm
{\it Acknowledgements.}
ST thanks Gerald Quinlan for many
discussions on this and related subjects.
WH thanks his supervisor, Prof.~Ken Jackson, for moral support for the
duration of this research, which was conducted during, but not related to,
his doctoral work.
We thank Rosemary McNaughton for many helpful comments on a draft of this
paper.
ST acknowledges support from an Imasco Fellowship.
This work was supported in part by the Natural Sciences and Engineering
Research Council of Canada, and the Information Technology Research Centre
of Ontario.

\appendix

\section*{Appendix: mean, variance, compression of uniform distributions}

If $n$ numbers are chosen from the uniform random interval $U[0,1]$
and sorted into non-decreasing order, then the mean and variance
of the $i^{\rm th}$ one $X_i$ are (Rice 1988, problem 4.15)\nocite{Rice88}
\begin{equation}
\bar X_i = i/(n+1),
    \;\;\;\;\;\sigma_{X_i}^2 = \frac{i(n-i+1)}{(n+1)^2 (n+2)}.
\label{eq:meanSigma}
\end{equation}
which we then scale to the interval $[\log(0.2), \log(50)]$ used in
this paper.  Radius exclusion makes the allowed intervals between
planets smaller, thus compressing the uniform standard deviations
(\ref{eq:meanSigma}) by a factor $C_i$.  To compute $C_i$, assume $r_i
= 1$ and that all the planets have the same fractional Hill radius
$H$ and perfectly fit a Bode's law with exponent $c=\beta$ and
$a=0$.  It is easy to show that
\begin{equation}
C_i = \frac{\log(\beta)}{\log(\beta) + \log(1-H) - \log(1+H)},
\label{eq:squishes}
\end{equation}
finally giving
\begin{equation}
\sigma_i^2 = (\sigma_{X_i}/C_i)^2,
\label{eq:sigma}
\end{equation}
which we substitute into equation (\ref{eq:objective}).  When distance
is measured in $\log r$ and all planets have the same radius exclusion,
equation (\ref{eq:squishes}) is exact; otherwise it is only approximate.

\bibliography{bode}


\end{document}